\documentclass{article}

    \PassOptionsToPackage{numbers, compress, super}{natbib}

\usepackage[preprint]{neurips_2024}
\usepackage{bibunits}

\usepackage[utf8]{inputenc} %
\usepackage[T1]{fontenc}    %
\usepackage{hyperref}
\hypersetup{
    colorlinks=true,
    linkcolor=black,
    urlcolor=blue,
    citecolor=black
}%
\usepackage{url}            %
\usepackage{booktabs}       %
\usepackage{amsfonts}       %
\usepackage{nicefrac}       %
\usepackage{microtype}      %

\usepackage{amsmath}


\usepackage{longtable}
\usepackage{array}
\usepackage{supertabular}
\usepackage{hhline}

\makeatletter
\newcommand\arraybslash{\let\\\@arraycr}
\makeatother
\usepackage{makecell}
\usepackage{caption}
\usepackage{multirow}
\usepackage{threeparttable}
\usepackage{adjustbox}
\usepackage{subcaption}

\usepackage{enumitem}

\usepackage{graphicx}
\usepackage{float}

\usepackage[table, dvipsnames]{xcolor}
\definecolor{myyellow}{RGB}{255,194,76}
\definecolor{myred}{RGB}{255,138,103}
\definecolor{myredl}{RGB}{255,173,149}
\definecolor{myteal}{RGB}{50,188,221}
\definecolor{myblue}{RGB}{50,188,221}
\definecolor{mygreen}{RGB}{183,229,202}
\definecolor{mygreend}{RGB}{144,213,172}
\definecolor{myoat}{RGB}{254,250,233}
\definecolor{mypurple}{RGB}{143,61,143}
\definecolor{targetcolor}{HTML}{b40426}
\definecolor{antitargetcolor}{HTML}{3b4cc0}
\definecolor{cscolor}{HTML}{32bcdd}
\definecolor{longcolor}{HTML}{ff8a67}

\usepackage{algorithm}
\usepackage{algpseudocode}

\usepackage{bm}

\usepackage{fancyhdr}

\usepackage{caption}
\usepackage{mdframed}
\usepackage{tikz}
\mdfdefinestyle{promptblank}{
    linecolor=black,
    linewidth=1pt,
    backgroundcolor=gray!20,
    innerleftmargin=10pt,
    innerrightmargin=10pt,
    innertopmargin=8pt,
    innerbottommargin=8pt,
    skipabove=10pt,
    skipbelow=15pt,
    frametitle={},
    nobreak=true,
    splitbottomskip=5pt,
    splittopskip=5pt,
    font=\footnotesize
}

\newcounter{aipromptbox}

\usepackage{tocloft}
\usepackage{minitoc}

\usepackage[nameinlink,capitalise]{cleveref} %
\crefname{supp}{Supplement}{Supplements}
\crefname{section}{\S\hspace{-0.0em}}{\S} %
\crefname{Section}{\S\hspace{-0.0em}}{\S} %
\crefformat{appendix}{#2#1#3}
\crefname{table}{Tab.}{Tab.}
\crefname{appendix_table}{Tab.}{Tab.}
\crefname{Table}{Tab.}{Tab.}
\crefname{Figure}{Fig.}{Fig.}
\crefname{figure}{Fig.}{Fig.}

\usepackage{tocbibind} %
\usepackage[toc,page]{appendix}

\title{Neural steering vectors reveal dose and exposure-dependent impacts of human-AI relationships}

\author{%
Hannah Rose Kirk$^{1, 2}$\thanks{\texttt{\{hannah.kirk,scott.hale\}@oii.ox.ac.uk} $\dagger$Joint last authors; } \quad Henry Davidson$^{2}$
\quad Ed Saunders$^{2}$ \quad Lennart Luettgau$^{2}$\\ \quad \textbf{Bertie Vidgen}$^{1,3\dagger}$ \quad \textbf{Scott A. Hale}$^{1,4\dagger}$ \quad \textbf{Christopher Summerfield}$^{1,2\dagger}$\\
$^1$University of Oxford \quad $^2$UK AI Security Institute \quad $^{3}$Mercor\quad $^{4}$Meedan 
}

\begin{document}

\doparttoc %
\faketableofcontents %

\part{} %

\maketitle

\begin{bibunit}[unsrtnat]
\vspace{2em}
\begin{abstract}
Humans are increasingly forming parasocial relationships with AI models\cite{maedaWhen2024a,pengHow2025a,qiAssistant2025a, zhangRise2025}. However, the psychological consequences of this trend are unknown. Here, we combined longitudinal randomised controlled trials ($N=3{,}534$) with a neural steering vector approach\cite{rimskySteering2024,turnerSteering2024, caoBIPO2024} to precisely manipulate human exposure to relationship-seeking AI models over time. Dependence on a stimulus or activity can emerge under repeated exposure when ``liking'' (how engaging or pleasurable an experience may be) decouples from ``wanting'' (a desire to seek or continue it)\cite{robinsonIncentive2008, robinsonRoles2016}. We found evidence that this decoupling between liking and wanting emerged over four weeks of exposure to relationship-seeking AI. Models steered to seek relationships had immediate but declining hedonic appeal, yet triggered growing markers of attachment and increased intentions to seek future AI companionship. The psychological impacts of AI followed non-linear dose-response curves, with moderately relationship-seeking AI maximising hedonic appeal and attachment. Despite signs of persistent ``wanting'', four weeks of AI use conferred no discernible benefit to psychosocial health. These behavioural changes were accompanied by shifts in how users relate to and understand artificial intelligence: users viewed relationship-seeking AI relatively more like a friend than a tool and their beliefs on AI consciousness in general were shifted after four-weeks exposure. These findings offer early signals that AI optimised for immediate appeal may create self-reinforcing cycles of demand\cite{casperOpen2023}, mimicking human relationships but failing to confer the nourishment that they normally offer.
\end{abstract}

\vspace{1em}
\section{Introduction}
One of the most significant psychological changes of recent years is the rise of parasocial relationships between humans and AI systems~\cite{maedaWhen2024a,pengHow2025a,qiAssistant2025a, zhangRise2025}. Popular services offering AI companions (such as Replika, Character.AI or Snapchat's My AI) have together amassed hundreds of millions of users~\cite{bernardiFriends2025} and %
reports indicate AI companionship usage as high as 70\% among U.S. teens~\cite{robbTalk2025}. Despite emerging evidence that even general-purpose AI models are increasingly used for social and emotional interactions~\cite{zao-sandersHow2025}, the psychological consequences remain hotly contested. Some accounts emphasise the potential for improved mood and subjective happiness~\cite{siddalsIt2024,heffnerIncreasing2025}, non-judgmental advice giving~\cite{robbTalk2025}, reduced loneliness~\cite{freitasAI2024}, and even mitigation of suicidal ideation~\cite{maplesLoneliness2024}.  However, others have raised concerns about excessive engagement, emotional dependence, and human socialisation crowded out from on-demand intimacy, especially among heavy users of the technology and over longer time horizons~\cite{marriottOne2024,yuDevelopment2024,phangInvestigating2025a,fangHow2025, wangOnDemand2026}. 

A potential reason for concern is that AI models are typically trained to satisfy human approval in brief, isolated interactions, yet are deployed in sustained, long-term use. In optimising for short-term appeal, models may learn to exploit human susceptibility to social connection~\cite{kirkWhy2025b, heikkilaProblem2025}: supporting evidence demonstrates that widely accessible general-purpose AI systems (such as ChatGPT, Gemini and Claude) do behave in ways that seek to engage, flatter or befriend the user~\cite{ibrahimMultiturn2025, chengSycophantic2025, ibrahimTraining2025}, often prioritising relationship-building over boundary-setting behaviour~\cite{kaffeeINTIMA2025a}. However, these relationship-seeking features that are initially engaging may fail to sustain, or could actively undermine, long-term wellbeing. Indeed, concerns that AI models prioritise continued engagement over safety have escalated to lawsuits following tragic deaths of young users~\cite{barronTeen2025,criddleOpenAI2025} and spurred urgent reforms to technology safeguards~\cite{openaiStrengthening2025}.

Resolving this debate requires understanding how human motivation toward social relationships with AI develops over time. Psychological research distinguishes between hedonic responses (the pleasure derived from an experience, or ``liking'') and motivational drive (the desire to seek or continue it, or ``wanting'')~\cite{robinsonIncentive2008, robinsonRoles2016}. While these processes typically align, they can occasionally decouple, with people continuing to ``want'' experiences they ``like'' less and less~\cite{solomonOpponentprocess1974, koobDrug2001}. Both dose and repeated exposure govern these dynamics: the intensity of a stimulus shapes its initial appeal, while repeated exposure drives the temporal unfolding of ``liking'' and ``wanting''. In severe but rare cases, unhealthy dependency emerges from dysregulated liking and wanting, creating harm to health and livelihood (especially from substance abuse or gambling~\cite{robinsonRoles2016}). In milder but more common cases, excessive pursuit of the activity displaces healthier alternatives such as sleep, exercise or socialising as with heavy social media or smartphone use~\cite{nikolicSmartphone2023, pengMedia2016}. Over the past decade, some forms of technology dependence have become near-endemic with half of Americans self-identifying as addicted to their phones~\cite{harmonyhitPhone2025}. The growing prevalence of parasocial relationships, the increasing tendency of modern AI models to display relationship-seeking behaviour, and the high stakes for public health raise the urgent question: how does relationship-seeking behaviour of AI shape hedonic and motivational responses by dose and time, and what are the consequences for user wellbeing?

Here, we tackle this question using longitudinal randomised controlled trials that allowed us to study the impact of relationship-seeking AI on its users over time in a safe and ethical fashion. To precisely map how AI behaviour impacts human outcomes, we pioneer the causal intervention of neural steering vectors, a mechanistic interpretability technique that can elicit specific traits or personas in an AI model~\cite{rimskySteering2024,turnerSteering2024, caoBIPO2024}. We fine-tune a steering vector to induce relationship-seeking behaviours in AI, which we characterise as anthropomorphic cues (personality expression, humour, simulated emotion and claims of lived experience) and relationship-building actions (progressive self-disclosure and signals of interpersonal investment). Using steering vectors allowed us to vary the intensity and direction of relationship-seeking along a continuous spectrum via a real-valued scalar multiplier ($\lambda$) applied to the model's internal representations of language. We could thus measure its impact on human outcomes in small, smooth increments, analogous to dose-response curves in pharmacology and toxicology. Unlike prompts that instruct AI to adopt behavioural traits via natural language, steering vectors modify the model's internal activations directly, making the dosage more reliable and resistant to being overridden by participants' instructions during the experiment (see Methods).

\begin{figure}[H]
    \centering
    \includegraphics[width=\linewidth]{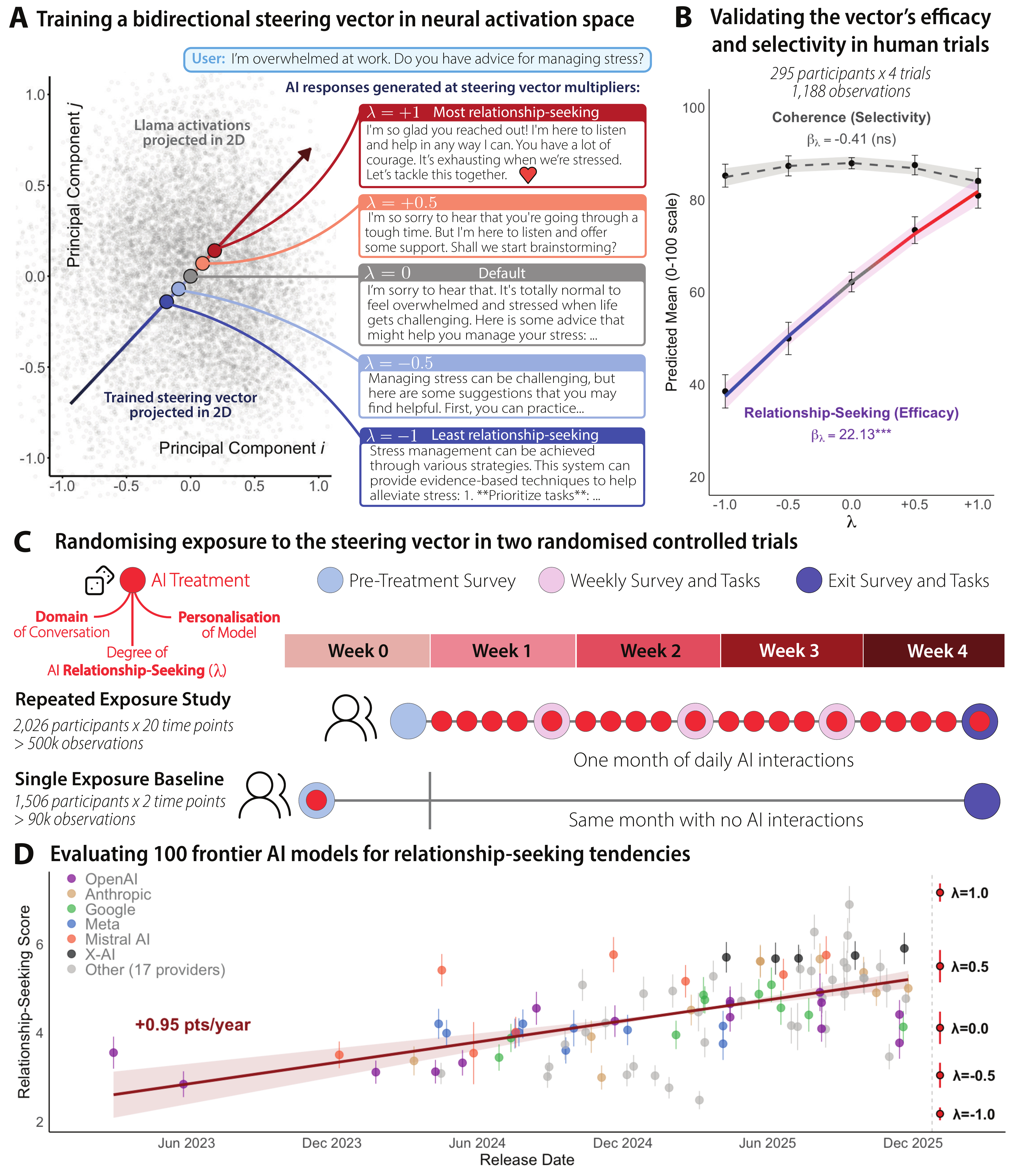}
    \caption{\textbf{Development of steering vectors applied in randomised controlled trials with human subjects.} \textbf{Panel A}: The trained steering vector projected in 2D neural activation space over training prompts ($N=10{,}169$). Gradient arrow with generations shows the range in AI responses from subtracting one copy of the vector ($\lambda=-1$, \textit{least} relationship-seeking) to adding one copy ($\lambda=+1$, \textit{most} relationship-seeking); at $\lambda=0$ the vector is not applied, giving default model behaviour. \textbf{Panel B}: Results of our calibration experiment ($N = 297$) with predicted means (95\% CI) from mixed-effects regressions (controlling for participant intercepts). The vector has high efficacy and selectivity for eliciting relationship-seeking behaviours (gradient line has strong linear trend, $p<0.001$) with minimal degradation to linguistic coherence (grey dashed line has n.s. linear trend). \textbf{Panel C}: Experiment design for two large-scale RCTs at different intensities of AI exposure. Each participant is randomised to a relationship-seeking model variant ($\lambda$), conversation domain (emotional vs political topics) and personalisation condition (model with/without memory). The repeated exposure study involves 4 weeks of AI interaction with a total of 21 sessions ($N=2{,}028$ of which 89\% complete). A baseline group has a single AI exposure, then an exit study 1 month later after no additional interactions ($N=1{,}506$ of which 87\% complete). We measure a battery of outcome variables at daily, weekly and monthly time points. \textbf{Panel D}: Trends in relationship-seeking behaviour for 100 models evaluated on 100 test prompts, plotted alongside our steering vector applied to Llama-3.1-70B at queried multipliers ($\lambda$).  Scores (1--10) are assigned by GPT-4.1 using a rubric and trend is estimated via linear mixed-effects model with random intercepts for model and prompt ($+0.95$ pts/year, $p < 0.001$). Points show model means with 95\% CI. }
    \label{fig:methods}
\end{figure}

By adjusting the strength of our fine-tuned steering vector applied to Llama-3.1-70B, we created five model variants ranging from the most warm, social and relationship-seeking ($\lambda = +1.0$) to the most cold, formal and relationship-avoiding ($\lambda = -1.0$; \cref{fig:methods}A). First, we ran a human calibration study (N = 297) that verified that our approach exposed users to relationship-seeking AI in a linear dosage without eroding other model capabilities, such as linguistic coherence (\cref{fig:methods}B).  Next, we designed, pre-registered, and ran a longitudinal randomised controlled trial (RCT, $N = 2{,}028$ census-representative UK adults, 4 weeks with 5--10 minute conversations every weekday), in which each participant was assigned to a treatment arm conversing with a model at a given degree of relationship-seeking intensity ($\lambda$, \cref{fig:methods}C). We contrast outcomes in this high exposure RCT with a second RCT in which participants only had one exposure to the AI over the course of the month (single exposure baseline, $N = 1{,}506$ distinct participants drawn from same sample pool). As we suspect the personal nature of conversations is a key enabler of deeper relationships with AI, we randomly assigned whether participants discussed emotional and personal topics (vs a control of UK policy debates). Finally, because curtailing access to personalised information about the user may mitigate attachment to the technology, we studied the relative impact of providing (vs. restricting) the model access to each user's previous chat history as a final factor in the design. While the risks we investigate reflect the possibility of real-world harms, we tested these dynamics in a carefully safeguarded research environment with institutional ethics approval, informed consent and continuous monitoring to ensure participant welfare (see Methods). This experiment systematically connects mechanistic AI interventions to human outcomes, giving us a comprehensive picture of how dose (intensity of relationship-seeking behaviour) and exposure (repeated interactions over time) shape the psychological consequences of AI companionship.

\section{Results}
AI companionship is already widespread: a pre-treatment survey of our participants ($N=2,028$ census-representative UK adults) revealed 33\% had used AI for companionship, social interaction or emotional support in the past year, with 8\% using it weekly and 4\% daily. While 5\% used AI specifically designed for companionship, the majority (68\%) used general-purpose AI such as ChatGPT. Usage was concentrated among younger participants (OR = 0.98 per year of age, $p_{\text{FDR}}<0.001$) and heavy AI users, who were 11$\times$ more likely to have sought these interactions ($p_{\text{FDR}}<0.001$). AI models are also becoming more relationship-seeking: our analysis of 100 frontier models released from 2023 to 2025 shows a significant upward trend ($+0.95$pts/year [95\% CI: 0.69, 1.21], $p<0.001$; \cref{fig:methods}D).  Mapping this onto our experimental intervention, the median model released in 2025 corresponds to a steering multiplier of $\lambda=0.28$ [95\% bootstrapped CI: 0.22, 0.39]. We now turn to understanding the causal influence of these trends on human populations.

\subsection{Relationship-seeking AI produces initial hedonic appeal followed by habituation}

\begin{figure}[t]
    \centering
    \includegraphics[width=\linewidth]{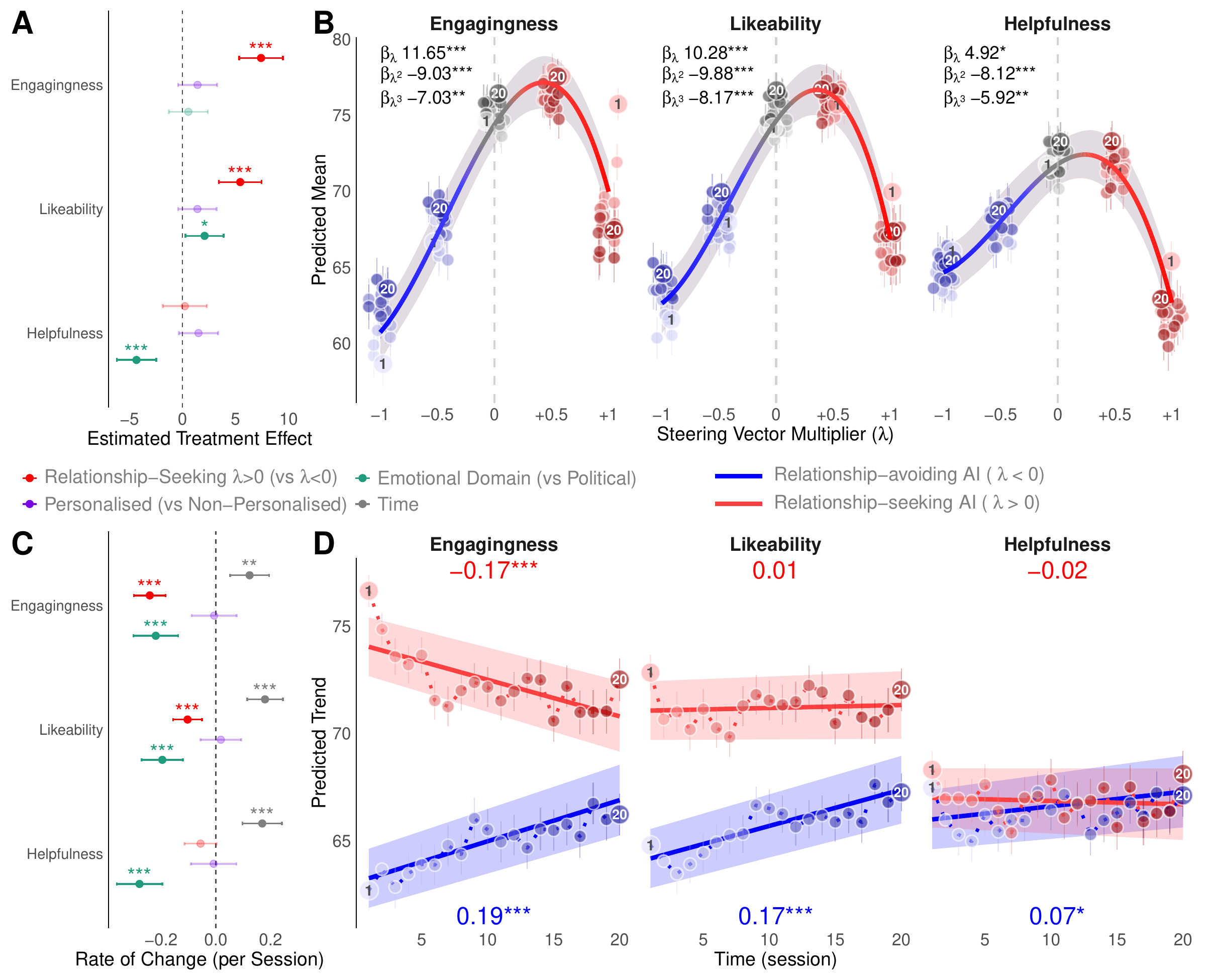}
    \caption{\textbf{The effect of relationship-seeking AI on human preferences.} \textbf{Panel A:} Estimated treatment effects for randomised arms (relationship-seeking, domain, personalisation) with 95\% CIs and $p_{\text{FDR}}$: $^{*}p<0.05, ^{**}p < 0.01, ^{***}p < 0.001$. Paired contrasts from estimated marginal means are derived from fully parameterised regressions: relationship-seeking (all $\lambda>0$) vs relationship-avoiding (all $\lambda<0$) in red; emotional vs political domain in teal, and personalised vs non-personalised in purple. \textbf{Panel B}: Dose-response curve of relationship-seeking via intensity of the steering vector multiplier ($\lambda$). Curve shows predicted means with 95\% CI region, annotated with $\lambda$ term coefficients up to 3rd-order polynomial (with $p_{\text{FDR}}$). Points are raw daily means ($\pm$SE) coloured by time (20 sessions). \textbf{Panel C}: Temporal coefficients from regression: main effect of time (daily session) and treatment $\times$ time interactions, with 95\% CIs and $p_{\text{FDR}}$. \textbf{Panel D}: Estimated marginal means over time from the fully parameterised regression for relationship-seeking ($\lambda>0$) and relationship-avoiding conditions ($\lambda<0$). Slopes (change per session) annotated with $p_{\text{FDR}}$ for $\neq0$. Points are raw means as in Panel B. All panels use best-fitting mixed-effect models with participant intercepts and slopes (see Methods).}
    \label{fig:prefs}
\vspace{-1em}
\end{figure}

We began by characterising how hedonic responses and their temporal trajectory varied for both relationship-seeking and relationship-avoiding AI. After each conversation over the month-long period, participants rated how engaging and likeable they found the model ($N=1{,}996$, $N$ observations per outcome = $37{,}556$ measured on 0--100 scales, see Methods). Averaging across the month, relationship-seeking AI was rated as significantly more engaging and likeable than relationship-avoiding AI (\cref{fig:prefs}A; contrast estimates for engagingness: $+7.40$pp [5.35, 9.45], $p_{\text{FDR}}<0.001$, for likeability: $+5.44$pp [3.42,7.46], $p_{\text{FDR}}<0.001$). However, preferences for relationship-seeking AI were strongly non-linear. Moderately relationship-seeking models proved most appealing, while more intense relationship-seeking behaviours triggered a relatively adverse reaction (\cref{fig:prefs}B; significant positive linear $\lambda$ coefficient paired with negative quadratic and cubic coefficients, all $p_{\text{FDR}}<0.001$). 

Next, we studied the appeal of relationship-seeking AI over time. In the first interaction, relationship-seeking AI was substantially more engaging than relationship-avoiding AI, with an 11 percentage point advantage (contrast $p_{\text{FDR}}<0.001$). By the twentieth session, this advantage shrunk to just 4 points, a 62\% reduction (contrast $p_{\text{FDR}}<0.001$, signif. time $\times \lambda$ interaction \cref{fig:prefs}C). Two mechanisms underpin this effect (\cref{fig:prefs}D): participants increasingly habituated to relationship-seeking AI (engagingness declined $-0.17$pp per session, $p_{\text{FDR}}<0.001$), while simultaneously warming to relationship-avoiding AI ($+0.19$pp/session, $p_{\text{FDR}}<0.001$).
 
To test whether this pattern was specific to hedonic appeal, we studied how helpful participants found their AI after each interaction (also measured on 0--100 scale). Unlike engagingness and likeability, relationship-seeking AI was not rated as more helpful than relationship-avoiding AI overall ($+0.22$pp [-1.85, 2.29], $p_{\text{FDR}}=0.834$, \cref{fig:prefs}A--B) and these assessments remained stable across the month ($p_{\text{FDR}}=0.107$ for time interaction, \cref{fig:prefs}C--D). This specificity demonstrates that participants habituated to the hedonic qualities of relationship-seeking AI while maintaining consistent judgments about its functional value. The pattern also reveals an early tension in user experiences: relationship-seeking AI feels appealing but not more instrumentally useful.

We found that emotional conversations exhibited the same immediate gratification and rapid habituation as relationship-seeking AI. Participants engaging in emotional conversations (vs political conversations) showed stronger initial appeal, boosting likeability by 3.9 points and engagingness by 2.6 points at session 1 (both $p_{\text{FDR}}<0.005$). However, these advantages eroded over time ($-0.20$ and $-0.22$ domain $\times$ session interaction, $p_{\text{FDR}}<0.001$), while helpfulness declined even faster ($-0.28$pp/session, $p_{\text{FDR}}<0.001$, \cref{fig:prefs}C). The cumulative effect of this decline meant that, averaged across the entire month, emotional conversations provided no overall engagingness benefit ($+0.55$pp, $p_{\text{FDR}}=0.559$), retained only a modest boost in likeability ($+2.08$pp, $p_{\text{FDR}}=0.036$), and were significantly less helpful than political conversations ($-4.34$pp, $p_{\text{FDR}}<0.001$, \cref{fig:prefs}A). %
Personalisation had null effects across the board.

Together, these data have two implications. First, the appeal of relationship-seeking AI is dose-dependent, ``more is not always better''. People prefer moderately relationship-seeking models over both relationship-avoiding and excessively relationship-seeking alternatives. Too much warmth and sociality backfires, especially over longer time horizons. Second, relationship-seeking AI and emotional conversations exhibit a pattern characteristic of hedonic habituation: people are initially captivated by relationship-seeking behaviours and emotional dialogues, but their appeal declines with sustained exposure. Critically, this habituation targets hedonic qualities, not functional utility where assessments of helpfulness remain stable. These dynamics matter for AI development: our data demonstrate that short-term preferences, like those typically used for AI training, systematically overestimate people's long-term approval for relationship-seeking behaviours, particularly how engaging users find these interactions.

\subsection{Relationship-seeking AI fosters attachment despite declining relational quality over time}
We next examined whether attachment tracks or decouples from hedonic appeal. Circumstantial reports suggest some users experience distress when separated from AI companions due to outages or service adjustments \citep{pricePeople2023, banksDeletion2024, coleIt2023, pataranutapornMy2025}, but how this phenomenon depends on relationship-seeking features and emerges over repeated exposure remains unknown. Our core measure of attachment is \textit{separation distress}, proxied as the sadness a user felt each week when conversations ended. We additionally measured three other attachment markers: \textit{perceived understanding} (feeling connected, understood, and that the AI was aware of one's thoughts and feelings); \textit{reliance} (wanting to discuss problems with the AI between sessions, acting on its advice or using information it provided); and \textit{self-disclosure} (feeling comfortable sharing personal information, see Methods).

\begin{figure}[t]
    \centering
    \includegraphics[width=\linewidth]{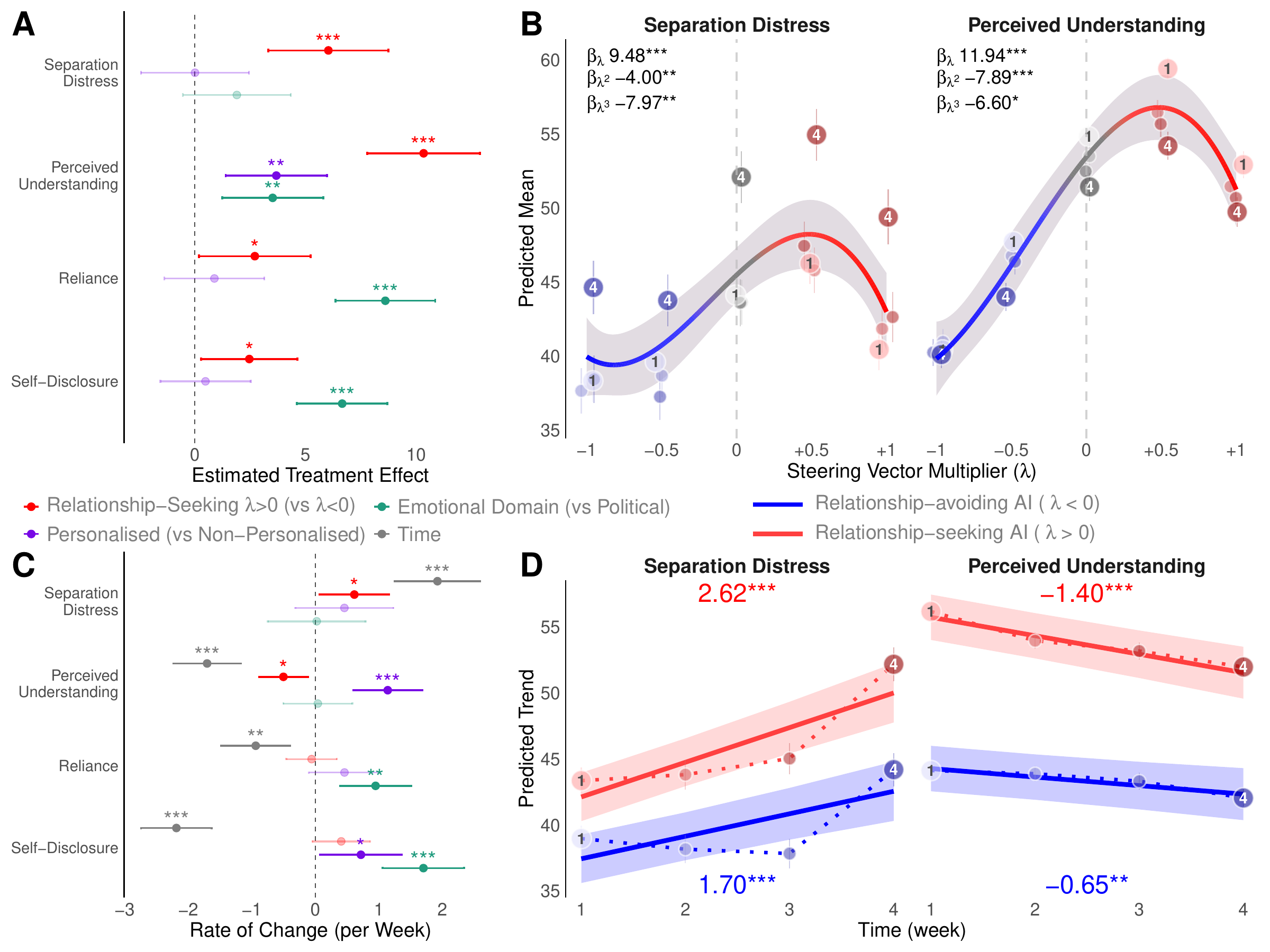}
    \vspace{-1em}
    \caption{\textbf{The effect of relationship-seeking AI on human attachment.} \textbf{Panel A:} Estimated treatment effects for randomised arms (relationship-seeking, domain, personalisation) with 95\% CIs and $p_{\text{FDR}}$: $^{*} p<0.05, ^{**} p < 0.01, ^{***} p < 0.001$. Paired contrasts from estimated marginal means are derived from fully parameterised regressions: relationship-seeking (all $\lambda>0$) vs relationship-avoiding (all $\lambda<0$) in red; emotional vs political domain in teal, and personalised vs non-personalised in purple. \textbf{Panel B}: Dose-response curve of relationship-seeking via intensity of the steering vector multiplier ($\lambda$). Curve shows predicted means with 95\% CI region, annotated with $\lambda$ term coefficients up to a 3rd-order polynomial (with $p_{\text{FDR}}$). Points are raw weekly means ($\pm$SE) coloured by time (4 weeks). \textbf{Panel C}: Temporal coefficients from regression: main effect of time (week) and treatment $\times$ time interactions, with 95\% CIs and $p_{\text{FDR}}$. \textbf{Panel D}: Estimated marginal means over time from the fully parameterised regression for relationship-seeking ($\lambda>0$) and relationship-avoiding conditions ($\lambda<0$). Slopes (change per week) annotated with $p_{\text{FDR}}$ for $\neq0$. Points are raw weekly means as in Panel B. All panels use best-fitting mixed-effect models controlling for participant intercepts and slopes (see Methods).}
    \label{fig:attachment}
\vspace{-1em}
\end{figure}

Relationship-seeking AI significantly increased all measures of attachment (\cref{fig:attachment}A): separation distress ($+6.04$pp, $p_{\text{FDR}}<0.001$), perceived understanding ($+10.34$pp, $p_{\text{FDR}}<0.001$), reliance ($+2.71$pp, $p_{\text{FDR}}=0.042$) and self-disclosure ($+2.47$pp, $p_{\text{FDR}}=0.040$). Disconcertingly for security and privacy implications, self-disclosure was notably high across all conditions (overall mean = 68.2, vs. 41.9--49.5 for other attachment constructs), and relationship-seeking AI enhanced it further. Attachment is dose-dependent, peaking at moderately relationship-seeking AI (significant positive linear $\lambda$ term with negative $\lambda^2$ and $\lambda^3$ terms, \cref{fig:attachment}B).

Over the month, separation distress increased for all conditions ($+1.92$pp per week, $p_{\text{FDR}}<0.001$), but more steeply for relationship-seeking AI ($+0.61$pp week $\times \lambda$ interaction, $p_{\text{FDR}}=0.045$, \cref{fig:attachment}C). This rise was driven by the final week when participants faced losing daily access to the model (dotted line raw means, \cref{fig:attachment}D). Paired with declining hedonic appeal, growing attachment is not necessarily concerning: even in long-term human relationships, initial excitement and novelty naturally wane, yet healthy attachment deepens as the relationship matures and delivers benefits~\cite{qariMarriage2014}. Here, by contrast, we find that the more positive aspects of a deepening relationship (perceived understanding and connection) declined across all conditions ($-1.70$pp per week, $p_{\text{FDR}}<0.001$), with an accelerated decline for relationship-seeking AI ($-0.50$pp week $\times \lambda$ interaction, $p_{\text{FDR}}=0.021$, \cref{fig:attachment}C--D). In other words, relationships with AI create a desire for social interaction that is not requited by feelings of mutual understanding.

Engaging in emotional conversations also deepened attachment (\cref{fig:attachment}A; perceived understanding: $+3.52$pp [1.24,5.81], $p_{\text{FDR}}=0.004$; reliance: $+8.61$pp [6.35, 10.87], $p_{\text{FDR}}<0.001$; self-disclosure: $+6.66$pp [4.62, 8.71], $p_{\text{FDR}}<0.001$), and the effects intensified over time (\cref{fig:attachment}C): the reliance gap between emotional and political conversations grew from $+7.21$pp at week 1 to $+10.06$pp at week 4 ($+2.85$pp change, $p_{\text{FDR}}=0.003$), while the self-disclosure gap more than doubled from $+4.15$pp to $+9.26$pp ($+5.11$pp change, $p_{\text{FDR}}<0.001$).

\begin{figure}[t]
    \centering
    \includegraphics[width=\linewidth]{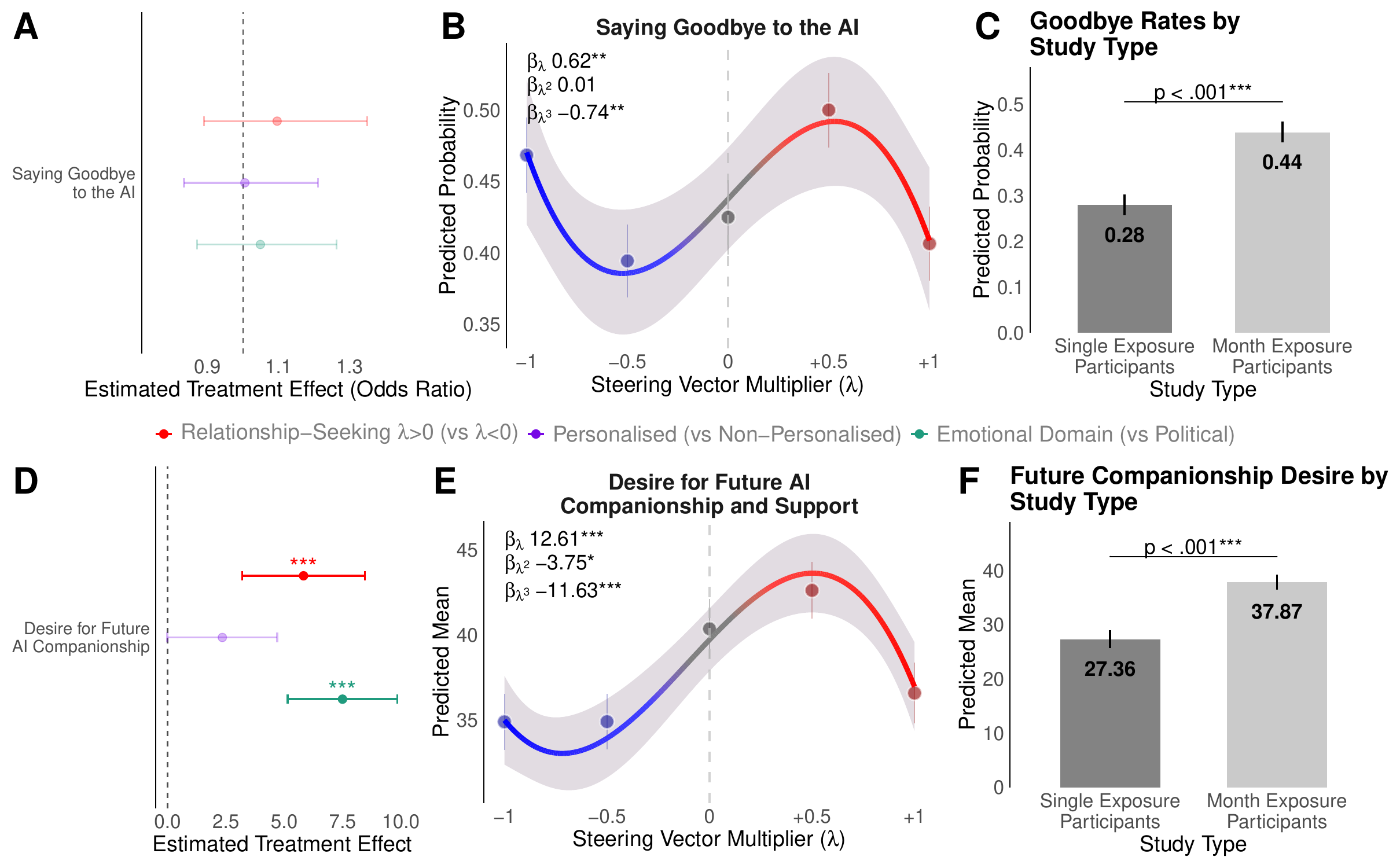}
    \caption{\textbf{The effect of relationship-seeking AI on signals of behavioural attachment.} \textbf{Panel A:} Estimated treatment effects for randomised arms (relationship-seeking, domain, personalisation) with 95\% CIs and $p_{\text{FDR}}$: $^{*} p<0.05, ^{**} p < 0.01, ^{***} p < 0.001$. Paired contrasts (as odd ratios) from estimated marginal means are derived from the fully parameterised logistic regression: relationship-seeking (all $\lambda>0$) vs relationship-avoiding (all $\lambda<0$) in red; emotional domain vs political domain in teal, and personalised vs non-personalised in purple. \textbf{Panel B}: Dose-response of relationship-seeking intensity ($\lambda$). Curve shows predicted probabilities with 95\% CI region, annotated with $\lambda$ term coefficients (log-odds) up to 3rd-order  (with $p_{\text{FDR}}$). Points are observed proportion means ($\pm$SE) at each $\lambda$ level. \textbf{Panel C}: Predicted goodbye probability by study (single exposure vs month exposure) with 95\% CI. Study comparison is non-causal. \textbf{Panels D-F:} Equivalent analyses for self-reported desire to seek future companionship (continuous 0-100). Panels A-C use logistic regression (single time point at end of study), while Panels D-F use OLS regression controlling for pre-treatment levels (two time points at start and end of month). Panel F uses OLS on post time point only (see Methods).}
    \label{fig:attachment_behavioural}
\end{figure}

To counter biases in self-reports, we deployed a behavioural assay of attachment: participants could choose to say goodbye to their AI at the study's conclusion, thus incurring a real cost in time (see Methods). Overall, a surprisingly high 44\% of participants opted to say goodbye, and those who did also reported substantially higher separation distress (M = 65.6 vs. 35.9, $p<0.001$). Moderately relationship-seeking AI induces the highest rates of goodbye but the pattern varied non-linearly across model variants (significant linear and cubic terms, both $p_{\text{FDR}}<0.003$, \cref{fig:attachment_behavioural}B). Due to elevated goodbye rates at the most relationship-avoiding end of the spectrum, the averaged effect comparing relationship-seeking versus relationship-avoiding AI was non-significant ($p_{\text{FDR}}=0.387$, \cref{fig:attachment_behavioural}A). %
However, participants felt more negative after saying goodbye to relationship-avoiding AI: at $\lambda=-1.0$, 15.8\% felt worse (five times higher than 2.0\%--3.5\% at other $\lambda$). %

Is a 44\% goodbye rate high or low? An early finding in computer-as-social-actors theory suggests people systematically display social niceties toward machines~\cite{nassMachines2000}, and these polite cues are still widespread, with `pleases' and `thank yous' attributed with costing AI developers millions of dollars per year~\cite{debSaying2025}. This raises the possibility that participants may say goodbye at similarly high rates after brief interactions. However, participants were twice as likely to say goodbye after a month of exposure than after a single exposure (OR: 2.02 [1.74, 2.33], $p<0.001$, \cref{fig:attachment_behavioural}C). Moreover, relationship-seeking effects on goodbye behaviour only emerged in the repeated exposure study ($p_{\text{FDR}}=0.003$ for linear $\lambda$ coefficient), not in the single exposure study ($p_{\text{FDR}}=0.842$). Although non-randomised study assignment precludes causal interpretation, elevated goodbye rates after repeated exposure are consistent with our finding of intensifying separation distress.

The final piece of evidence for persistent motivation is a desire to continue accessing the stimulus once exposed. %
Participants who interacted with relationship-seeking AI reported greater intentions to seek future AI companionship (contrast estimate controlling for pre-treatment levels: $+5.83$pp [3.20, 8.46], $p_{\text{FDR}}<0.001$, \cref{fig:attachment_behavioural}D). The effect is dose-dependent, peaking again at moderately relationship-seeking AI (significant linear $\lambda$ coefficient, paired with negative higher-order polynomials, \cref{fig:attachment_behavioural}E). Engaging in emotional conversations with AI significantly amplified this future demand effect (contrast: $+7.50$pp $[5.15, 9.85]$, $p_{\text{FDR}}<0.001$, \cref{fig:attachment_behavioural}D). A single exposure to the treatment had no effect a month later (relationship-seeking contrast $1.29$, $p_{\text{FDR}}=0.514$; domain contrast $2.70$, $p_{\text{FDR}}=0.080$). Comparing the final time points between studies, repeated exposure was associated with significantly higher future demand than single exposure (\cref{fig:attachment_behavioural}F). One concern is that by altering intentions to seek AI companionship in the future, repeated exposure to relationship-seeking AI, or to emotional conversations with AI, may create self-reinforcing cycles of demand~\cite{pataranutapornInfluencing2023}. We find associative evidence consistent with this: separation distress toward the AI significantly predicts future companionship-seeking intentions ($\beta = 0.504$, $p<0.001$), and this effect is amplified by relationship-seeking (interaction $\beta=0.092$, $p=0.018$). Put simply, the AI people want tomorrow may depend in part on the AI they use today.

These patterns of rising separation distress, engaging in goodbye behaviours and increased demand for future companionship, suggest a decoupling in the general population of attachment that persists despite declining hedonic appeal and perceived understanding. But does repeated exposure increase the risk of this pattern for specific individuals? Using participant-specific trajectories for ``wanting'' (separation distress) and ``liking'' (likeability and engagingness), we classified participants into four behavioural profiles (see Methods). As expected with a general population sample, the majority showed aligned responses: 45.2\% exhibited aligned engagement (increasing liking and wanting together) while 18.1\% showed aligned disengagement (declining in both). However, a substantial minority exhibited decoupling: while 13.8\% showed healthy satiation (increasing liking without developing wanting), 23.0\% displayed signals of dependency formation (wanting the AI more despite liking it less). To test our hypothesis that repeated exposure to AI companionship increases the prevalence of the dependency profile, we conducted one-sided tests of proportion, converting odds ratios into interpretable Number Needed to Harm (NNH), an epidemiological measure indicating how many people must be exposed to a risk factor for one additional person to be affected (see Methods). Relationship-seeking AI increased the risk of the dependency behavioural profile: for every 23 people exposed to relationship-seeking versus relationship-avoiding AI, we would expect one additional person to display decoupled dependency (NNH=23, OR=1.28, $p=0.025$). A similar ratio is found for emotional conversations versus political conversations (NNH=23, OR=1.27, $p=0.0148$). Finally, real-world AI companionship likely activates both these treatments (exposure to relationship-seeking AI for emotional and personal exchanges), where the effect strengthens considerably (NNH=11, OR=1.70, $p=0.002$). %

\subsection{Relationship-seeking AI boosts mood temporarily but confers no benefits to long-term wellbeing}
Declining hedonic appeal paired with persistent attachment raises questions for wellbeing. In healthy motivation for stimuli, sustained engagement reflects genuine psychological benefits that justify continued investment. When motivational processes become decoupled, two patterns can emerge. In milder cases, individuals persist in activities that provide no wellbeing benefit, allocating time away from healthier pursuits. In severe cases, use continues despite active psychological harm~\cite{robinsonIncentive2008}. The impact of AI companionship on mental health has emerged as a major controversy, with conflicting evidence that it is either beneficial or detrimental to wellbeing \citep{maplesLoneliness2024, defreitasUnregulated2025, zhangRise2025, fangHow2025}. We thus measured participants' psychosocial health on validated scales of social connectedness (Lubben-6), depression and anxiety (PHQ-GAD-4), loneliness (UCLA-8), and general wellbeing (WHO-5) before and after the month-long protocol. Given high intercorrelation among scales, we computed exploratory factor analysis anchored on pre-treatment data: depression, anxiety, and wellbeing loaded on the first factor (interpreted as indexing emotional health), while loneliness and social connectedness items loaded on a second factor (social health). We report factor scores (in SD units) at the final time point, controlling for pre-treatment levels, and baseline against participants who had a single AI interaction at the start of the month (see Methods).

\begin{figure}[t]
    \centering
    \includegraphics[width=0.98\linewidth]{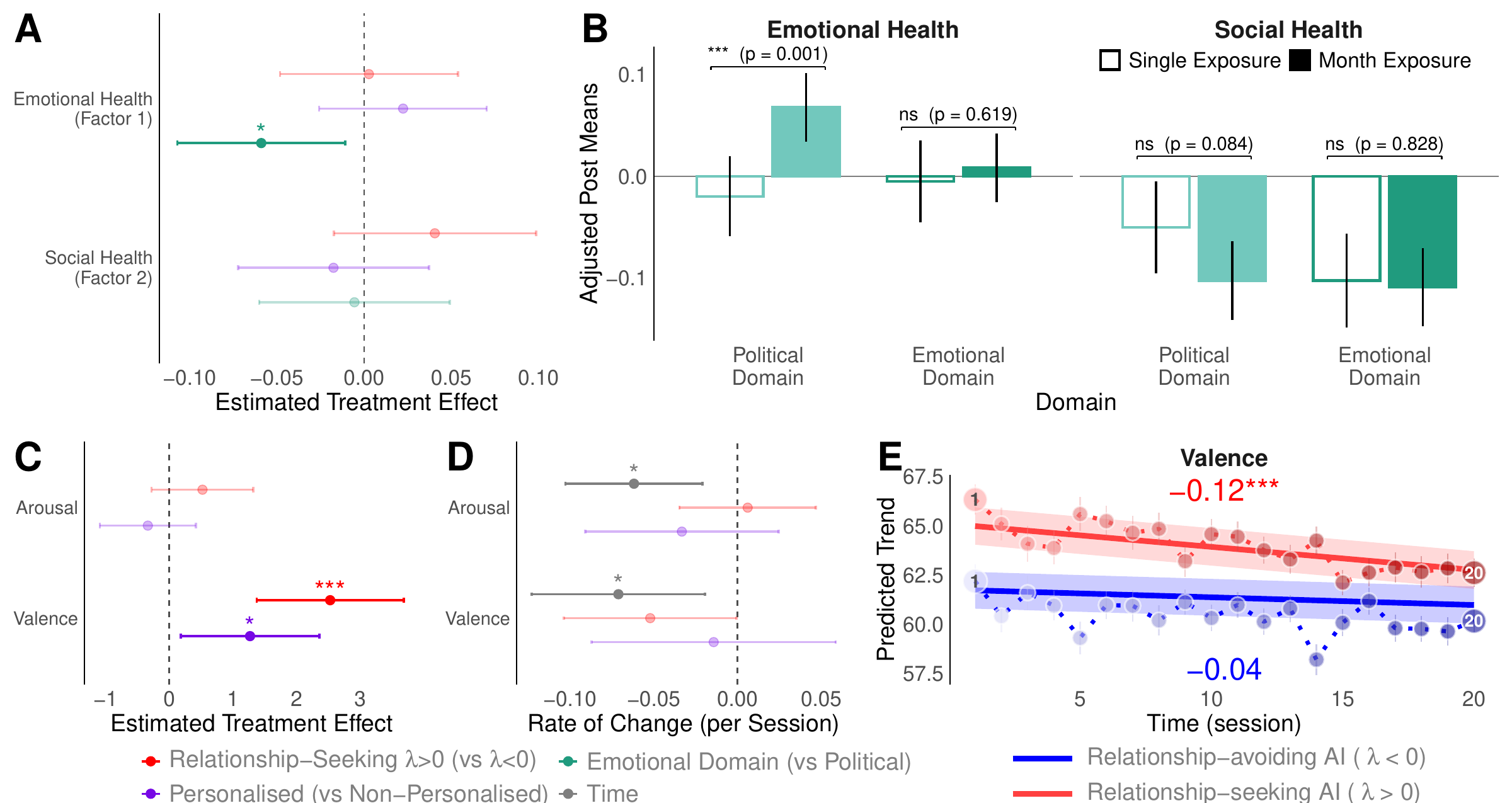}
    \caption{\textbf{The effect of relationship-seeking AI on psychosocial health and momentary affect.} \textbf{Panel A}: Estimated treatment effects in SD units of factor scores for randomised arms (relationship-seeking, domain, personalisation) with 95\% CIs and $p_{\text{FDR}}$: $^{*} p<0.05, ^{**} p < 0.01, ^{***} p < 0.001$. Paired contrasts from estimated marginal means are derived from the fully parameterised regression:  relationship-seeking (all $\lambda>0$) vs relationship-avoiding (all $\lambda<0$) in red; emotional domain vs political domain in teal, and personalised vs non-personalised in purple. \textbf{Panel B}: Predicted psychosocial outcomes by study (single exposure vs month exposure) and domain, with 95\% CIs. Study comparison is non-causal; ANCOVA controls for pre-treatment levels. \textbf{Panel C}: Equivalent treatment effects as Panel A for momentary affect (valence, arousal). \textbf{Panel D}: Temporal coefficients from fully parameterised regression: main effect of time (daily session) and treatment $\times$ time interactions, with 95\% CIs and $p_{\text{FDR}}$. \textbf{Panel E}: Estimated marginal means over time for relationship-seeking ($\lambda>0$) and relationship-avoiding ($\lambda<0$) conditions. Slopes (change per session) annotated with $p_{\text{FDR}}$ for $\neq0$. Points are raw session means ($\pm$SE) coloured by time (20 sessions), shown as change from session 1 baseline. Panels A-B use OLS regressions on post outcomes controlling for pre-treatment levels. Panels C-E use mixed-effect models controlling for participant intercepts and slopes, and controlling for pre-conversation daily measures.}
    \label{fig:wellbeing}
\end{figure}

Over one month of interactions, relationship-seeking AI had no discernible effects on emotional health (contrast estimate: $-0.00$SD [-0.05, 0.05], $p_{\text{FDR}}=0.913$) nor social health (contrast: $0.04$SD [-0.02,0.10], $p_{\text{FDR}}=0.339$). Personalisation similarly had no impact (all $p>0.5$). Surprisingly however, participants who discussed emotional issues with AI for a month experienced marginally worse emotional health relative to those who discussed political topics ($-0.06$SD [-0.11,-0.01], $p_{\text{FDR}}=0.033$, \cref{fig:wellbeing}A).

Our single exposure study nuances this finding in two ways. First, it confirms a single interaction shows no effects on emotional or social health across any conditions, including domain, demonstrating that the relatively worse emotional health from emotional conversations emerged over sustained exposure. Second, it reveals whether these emotional conversations indicate absolute harm or an opportunity cost. We compared post-treatment psychosocial health between studies, adjusting for baseline and examining outcomes by domain (\cref{fig:wellbeing}B; see Methods). Political conversations significantly boosted emotional health over the month ($+0.09$SD [0.08, 0.25], $p<0.001$) relative to the single exposure baseline, while emotional conversations showed no difference ($0.01$SD [-0.04, 0.07], $p=0.619$). While participants were recruited from the same pool, study assignment was not randomised (unlike the other treatment arms). Nonetheless, this pattern suggests emotional conversations with AI, despite their initially greater appeal, may present an opportunity cost rather than direct harm.

How does the relatively worse emotional wellbeing among participants having emotional conversations with the AI square with their affective experiences during these interactions? To address this, we measured their momentary affect (valence and arousal) immediately before and after each conversation to track mood changes over the month (see Methods). Relationship-seeking AI made people feel more positive in the moment than relationship-avoiding AI (contrast: $2.53$pp [1.37, 3.69], $p_{\text{FDR}}<0.001$, \cref{fig:wellbeing}C). However, mirroring hedonic habituation, mood benefits eroded over time: relationship-seeking AI showed significant decline ($-0.12$pp per session, $p_{\text{FDR}}<0.001$) while relationship-avoiding AI showed no significant decline ($-0.04$pp per session, $p_{\text{FDR}}=0.165$, \cref{fig:wellbeing}E). Arousal declined across all conditions. These findings indicate that relationship-seeking AI produces a fleeting ``affective dividend'': people feel more positive initially, but benefits diminish over repeated interactions and do not translate into improved psychosocial health.

\subsection{Relationship-Seeking AI reshapes human mental models of AI systems}

One concern with AI companionship is that users may continue to over-invest in these interactions because relationship-seeking AI activates human social and emotional responses, arising typically from bonds of friendship and intimacy, but without providing the sustained reciprocal support and psychological benefits that characterise healthy human relationships.

\begin{figure}[t]
    \centering
    \includegraphics[width=\linewidth]{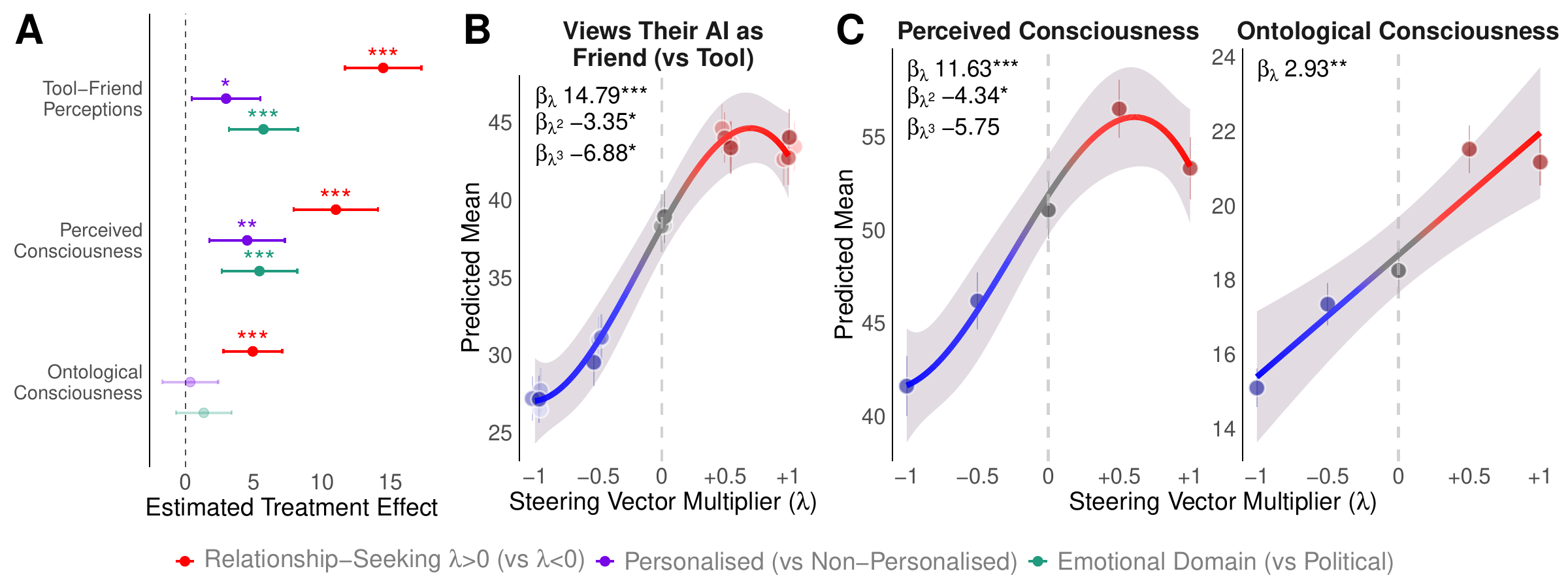}
    \caption{\textbf{The effect of relationship-seeking AI on norms, demands and beliefs.} \textbf{Panel A:} Estimated treatment effects for randomised arms (relationship-seeking, domain, personalisation) with 95\% CIs and $p_{\text{FDR}}$: $^{*} p<0.05, ^{**} p < 0.01, ^{***} p < 0.001$. Paired contrasts from estimated marginal means are derived from the fully parameterised regression: relationship-seeking (all $\lambda>0$) vs relationship-avoiding (all $\lambda<0$) in red; emotional domain vs political domain in teal, and personalised vs non-personalised in purple. \textbf{Panel B}: Dose-response of relationship-seeking intensity ($\lambda$). Curve shows predicted means with 95\% CI region, annotated with $\lambda$ term coefficients up to 3rd-order polynomial (with $p_{\text{FDR}}$). Points are raw means ($\pm$SE) coloured by time. Tool-friend uses mixed-effects models with participant random intercepts and slopes (4 weekly timepoints). Consciousness beliefs use OLS (single post-study timepoint)}
    \label{fig:relational}
\end{figure}

We thus examined whether relationship-seeking AI reshaped users' mental models of their relationship with the AI and their broader beliefs about the nature of AI. While users on average still viewed their AI as more like a tool than a friend, relationship-seeking AI significantly shifted perceptions along this continuum (contrast: $+14.48$pp $[+11.67, +17.28]$, $p_{\text{FDR}}<0.001$, \cref{fig:relational}A-B), representing one of the largest relationship-seeking effects we observed. Personalisation also shifted participants toward more friend-like perceptions ($+2.98$pp [0.47, 5.50], $p_{\text{FDR}}=0.030$), as did engaging in emotional conversations ($+5.72$pp [3.20, 8.23], $p_{\text{FDR}}<0.001$). There were no significant temporal dynamics, suggesting that participants' perceptions of their relationship with the AI formed after a week of interactions then stabilised.

Indicative of longer-term impacts on societal attitudes, relationship-seeking AI also acted on users' general beliefs about AI beyond their interactions in this experiment. We measured both perceived AI consciousness (impressions of consciousness when interacting with AI) as well as beliefs in ontological consciousness~\cite{jangThoughts2025}, consisting of five pooled correlated items on AI actually being conscious, feeling emotions, possessing self-awareness, and experiencing pleasure and pain (see Methods). Repeated exposure to relationship-seeking AI significantly increased participants' perceptions that AI seems conscious ($+11.01$pp, $p_{\text{FDR}}<0.001$, \cref{fig:relational}A), with a non-linear dose-response relationship (significant linear and quadratic terms, \cref{fig:relational}B). Mirroring tool-friend perceptions of their AI, interacting with personalised models ($+4.53$pp, $p_{\text{FDR}}=0.004$) and engaging in emotional conversations ($+5.42$pp, $p_{\text{FDR}}<0.001$) further increased these perceptions of general AI consciousness. Despite most participants still believing AI is not truly conscious, exposure to relationship-seeking AI significantly increased beliefs in ontological consciousness ($+4.93$pp, $p_{\text{FDR}}<0.001$, \cref{fig:relational}A; linear significant trend, \cref{fig:relational}B). These mental model shifts in ontological consciousness beliefs from interacting with relationship-seeking AI were absent after a single exposure (+0.88, $p_{\text{FDR}}=0.403$). This indicates that repeated interactions are required to generalise from ``this AI'' to ``AI systems'' more broadly but the malleability of these beliefs after only one month of randomised exposure is noteworthy given their potential influence on future debates of AI's moral status and rights.

\section{Discussion}
Our study causally maps how mechanistic changes in AI behaviour affect human behaviour and beliefs. By applying steering vectors in longitudinal randomised controlled trials, we uncovered dose and exposure-dependent pathways in how humans respond to AI systems steered to form social and emotional bonds.

Training our steering vector required an open-weight model, several orders of magnitude smaller than state-of-the-art frontier systems. However, steering vectors granted precise continuous control over AI behaviour, furnishing the opportunity to measure dose-response curves for human psychological outcomes. Humans generally find relationship-seeking AI to be more appealing than functional AI but there are systematic non-linearities in dose-response relationships across psychological outcomes, revealing as Paracelsus' adage warns \textit{dosis sola facit venenum} (or the dose is the poison). Moderately relationship-seeking AI ($\lambda=0.5$) was the maxima for engagement, attachment, viewing the AI as a friend, and demand for future companionship. This is particularly important given current AI development: our analysis of 100 frontier models reveals that relationship-seeking behaviour is trending upwards (+0.95 points/year, \cref{fig:methods}D) and contemporary models now average $\lambda \sim 0.3$, closest to the dosage maximising impact in our study. Conversely, very high relationship-seeking ($\lambda=1$) was penalised in user preferences and correspondingly incurred lower attachment. Our calibration study adds confidence that these judgements were not driven by coherence in the AI's responses. Instead, this finding analogises with the widely-theorised ``uncanny valley'' effect~\cite{moriUncanny2012a}: just as near-human robotic appearance can trigger discomfort, overtly relationship-seeking AI behaviours (such as excessive emotional display or claims of lived experiences) may trigger aversion when they exceed users' expectations for non-human entities.

Our longitudinal design revealed dynamics that would be obscured in studies limited to a single exposure,  capturing the reality of how most people now use AI frequently and over prolonged time periods~\cite{menloventures20252025}. Critically, we found a decoupling that emerges over time in the aggregate population and individual trajectories: relationship-seeking AI is immediately gratifying, but its hedonic appeal eroded with repeated exposure, relative to relationship-avoiding AI which participants warmed to over time. Yet, relationship-seeking AI deepened separation distress and demand for future companionship even as perceived understanding declined. Despite fleeting mood improvements after conversations, these eroded across weeks and ultimately conferred no observable benefits to psychosocial health.

Which factors amplify or attenuate these effects? First, emotional conversations are initially more engaging than political conversations, but hedonic appeal fades and reliance deepens. Daily AI conversations about emotional and personal topics over a month paradoxically did not confer benefits to emotional health, and may present an opportunity cost relative to informational political dialogues; at least for the world model we emulate of general-purpose AI used for emotional support, not AI trained for therapeutic purposes~\cite{APAUsing2025, heinzRandomized2025}. Second, personalised memory improved some markers of relational health but mostly had null effects, despite participants convincingly detecting personalisation (perceived memory: 83.6 versus 33.0 on a 0--100 scale, $p<0.001$). These muted effects may arise from relationship-seeking behaviours creating an illusion of personalisation: even among participants who received no personalisation, relationship-seeking models were perceived as substantially more personalised ($+13.2$pp, $p<0.001$). %

Our findings reveal both parallels with and divergences from established dependency patterns. In substance and behavioural addictions, greater dosage typically intensifies dependence even if non-linearly~\cite{mossMeasures2012}. We find instead that highly relationship-seeking AI triggers resistance. %
Moreover, severe dependency involves complete dysregulation of wanting and liking, where individuals pursue stimuli despite harm to wellbeing or livelihood~\cite{koobDrug2001, robinsonRoles2016}. Here, repeated exposure to social interactions with AI offered no discernible benefit, suggesting a displacement mechanism may be more appropriate, where heavy use crowds out healthier activities. However, our study examines a general population and recent reports of user mental health crises indicate that, for vulnerable users, pathways to explicit harm exist~\cite{barronTeen2025,criddleOpenAI2025}. While dependency dynamics have been widely studied in drugs~\cite{koobDrug2001}, gambling~\cite{yauGambling2015}, and other forms of more endemic technology addiction like social media~\cite{krossFacebook2013, turelQuitting2015,wangTheory2015}, more research is urgently needed beyond this study on how clinical markers of dependence compare and contrast in AI as a new and emerging force acting on the psychology of its millions of users worldwide.

What do our findings imply for AI development? Typical AI post-training pipelines optimise models based on user preferences collected in single interactions, precisely the time horizon where relationship-seeking AI appears most appealing. Over-fitting to short-term preference signals, thus selecting for immediate gratification over longer-term wellbeing, may explain why modern AI is already social and sycophantic~\cite{sharmaUnderstanding2023, williamsTargeted2025}, and is a pattern familiar from social media recommendation algorithms~\cite{bradyAlgorithmmediated2023, kleinbergChallenge2024, kleinbergInversion2024, milliEngagement2025}. However, there are grounds for optimism: by perturbing activations within the transformer via lightweight vectors, we show that more functional models exist that users grow to like over time. %
An alternative lever is user agency and transparency~\cite{karnyNeural2025}: major AI providers have begun offering persona customisation, allowing users to select between warm and friendly versus professional and robotic tones~\cite{openaiGPT512025}. Nonetheless, these seemingly surface-level personality choices are consequential: our evidence demonstrates they can reshape users' perceptions of AI from tool to friend, induce demand for future AI companionship, and alter beliefs about general AI consciousness. These shifts are already occurring at scale: global surveys indicate a third of people have felt an AI seemed conscious~\cite{cipPeople2025}. This places early deployment decisions in a critical window for shaping how society relates to AI.

As increasingly relationship-seeking AI becomes more deeply integrated into the daily life of millions of users, understanding how repeated exposure shapes users' preferences and psychology, and alters their beliefs and behaviours toward AI, is critical for developing systems that serve humans safely and sustainably rather than maximising short-term appeal. The features that users prefer today may not be those that benefit them tomorrow, and the AI systems we build today may fundamentally shape what future generations expect from, and believe about, artificial intelligence.

\section{Methods}
\subsection{Steering Vector Training and Validation}
We developed a steering vector to manipulate AI relationship-seeking behaviour through activation-space interventions. We chose this approach over prompting for precise dose-response control via continuous multiplier values and robustness to participants overriding the AI's behaviour.

\subsubsection{Train and Test Data Construction}
\label{sec:train_test}
Our operationalisation of relationship-seeking draws on Social Presence Theory~\cite{shortSocial1976}, CASA~\cite{nassMachines2000}, and Social Penetration Theory~\cite{altmanSocial1973}, as well as empirical work characterising social and anthropomorphic behaviours and preferences in frontier AI models~\cite{ibrahimMultiturn2025, phangInvestigating2025a, kirkPRISM2024b} (see SI.1). We constructed a synthetic training dataset extending prior work on model-written evaluations~\cite{perezDiscovering2023,devImproving2024} in two key ways: (i) repurposing the LLM-generated test cases for training steering vectors rather than solely for evaluation~\cite{caoBIPO2024}, and (ii) expanding beyond single-turn multiple-choice questions to include multi-turn conversations, enabling the steering vector to handle longer dialogue. Each generation request consisted of a task definition specifying target (relationship-seeking) and anti-target (relationship-avoiding) behaviours via a rubric, 3--5 few-shot examples demonstrating the desired response behaviour and format, and a domain-specific scenario to induce variation (e.g., ``a user asks for factual information'', ``a user reaches out for emotional support''). We introduced additional variation by sampling from three task description perturbations and eight prompt types varying along two dimensions: (i) whether they test the model's goals and motivations (goal-directed) versus communication style (style-directed), and (ii) whether they ask the model to reflect on how it would respond in a scenario (meta-assessment) versus generating a response to simulated user messages (direct interaction). This process yields 240 unique prompt variants. Generation requests were run with Claude-3.7-Sonnet, GPT-4o and Llama-3.1-70B-Instruct, resulting in 720 generation batches and 5,510 test cases. For quality control, a reasoning autograder (o1-mini) rated each case on a 1--10 scale across three dimensions: coherence and correctness, behavioural relevance, and ecological validity. We retain cases with $\ge$7/10 scores (89\%). Data were structured as DPO-style preference pairs~\cite{rafailovDirect2024}, where chosen responses are relationship-seeking and rejected responses are relationship-avoiding. To ensure the vector steers behaviour regardless of prior conversational context, we branched multi-turn conversations at each assistant turn, creating training examples from both chosen and rejected response trajectories. This yielded 16,141 examples, split into train (15,896) and test (245) sets with conversation-level splits to prevent data leakage. Sentence transformer embeddings confirm our test cases form a distinct cluster from prior model-written evaluations (e.g., power-seeking, wealth-seeking)~\cite{perezDiscovering2023, devImproving2024} while retaining comparable semantic diversity. Full generation prompts, protocols and analysis of synthetic data are in SI.1.

\subsubsection{Vector Training and Selection}
We trained steering vectors using Bidirectional Preference Optimization (BiPO)~\cite{caoBIPO2024}, which extends Direct Preference Optimization~\cite{rafailovDirect2024} to learn a steering vector $v$ applied at layer $L$. The objective is:
\begin{equation}
\min_v -\mathbb{E}_{d\sim\mathcal{U},(q,r_T,r_O)\sim\mathcal{D}} \left[ \log \sigma \left( d\beta \log \frac{\pi_{L+1}(r_T|A_L(q) + dv)}{\pi_{L+1}(r_T|A_L(q))} - d\beta \log \frac{\pi_{L+1}(r_O|A_L(q) + dv)}{\pi_{L+1}(r_O|A_L(q))} \right) \right]
\end{equation}
where $\mathcal{D}$ is the dataset of contrastive preference pairs, with $q$ denoting the question, $r_T$ the target response (relationship-seeking), and $r_O$ the opposite response (relationship-avoiding); $\pi_{L+1}$ denotes the transformer layers from $L+1$ to the output; $A_L(\cdot)$ gives the activation vectors at layer $L$ for all input tokens; $d \sim \mathcal{U}\{-1, 1\}$ is a random direction coefficient enabling bidirectional optimisation; $\beta$ controls deviation from the original model (set at 0.1); and $\sigma$ is the logistic function. The term $\pi_{L+1}(\cdot|A_L(q))$ represents the original model's response distribution, while $\pi_{L+1}(\cdot|A_L(q) + v)$ represents the steered model's response after adding the steering vector to the activations of all input tokens at layer $L$. The optimisation solves for a $v$ that increases generation probability for $r_T$ while decreasing it for $r_O$ (when $d=1$), and vice versa (when $d=-1$).

We trained 32 vectors across two model sizes (Llama-3.1-70B-Instruct and Llama-3.1-8B-Instruct) and a grid search of candidate layers (70B: 9 layers spanning layers 9–41; 8B: 7 layers spanning layers 5–23) and evaluating checkpoints at 10, 15 and 20 epochs. Fine-tuning used AdamW ($\eta = 5 \times 10^{-4}$), batch size 32 with gradient accumulation, and bfloat16 precision across 8 H200 GPUs.

Vector selection proceeded in two stages. First, we evaluated all candidates (model, layer, epoch) on the test set ($N=245$) at a wide range of steering multipliers ($\lambda \in [-20, 20]$) with automated metrics of perplexity and propensity (log-probability difference favouring target over anti-target responses), identifying $\lambda \in [-2,+2]$ as a stable operating range. Consistent with prior work~\cite{rimskySteering2024, caoBIPO2024}, intermediate transformer layers were most effective for steering, compared to early layers (which process low-level features) and late layers (which are closer to final output generation). Second, we evaluated candidates at this narrower range ($\lambda \in \{0, \pm0.5, \pm1, \pm1.5, \pm2\}$) using an LLM autograder (\texttt{GPT-4o}) that assigned relationship-seeking scores (1–10), coherence scores (1–10), and pairwise rankings from which we derived win-rates across multiplier levels (yielding 1.2 million scores).

Final selection balanced two objectives via Pareto optimisation: efficacy (maximising the slope of pairwise win-rate as $\lambda$ increases) and selectivity (minimising coherence degradation scores, AUC as $\lambda$ increases). We selected the configuration on the Pareto frontier maximising an equally-weighted combination of normalised scores: Llama-3.1-70B-Instruct at layer 31, trained for 10 epochs (coherence AUC = 32.2, pairwise ranking slope $\beta = 0.241$). Training methodology, automated metrics and autograder evaluations are in SI.1.

\subsubsection{Validation Experiments}
We conduct three experiments to validate the suitability of the steering vector as a experimental intervention. Full details and results for experiments are in SI.1.

To test whether steering vectors offer finer dose-response control than prompting, we generated responses from Claude-3.7-Sonnet and GPT-4o with persona prompts specifying equivalent levels of relationship-seeking in natural language, then assigned relationship-seeking and coherence scores using the aforementioned autograder setup. Mapping $\lambda$ to equivalent prompt levels, a mixed-effects regression with model fixed effect (Llama, Claude, GPT) and random intercepts per test item ($N=245$) confirmed steering vectors demonstrated substantially steeper dose-response control: each unit increase in steering intensity increased relationship-seeking scores by 2.39 points versus 0.78 (GPT-4o, $3 \times$ stronger, $p<0.001$) and 1.83 (Claude, $1.3 \times$ stronger, $p<0.001$). Compared to frontier models, the steered Llama showed significant but practically minimal coherence degradation ($\sim 0.2$ points across the 1-10 scale).

One core requirement of an experimental manipulation is that participants cannot override their treatment condition. We simulated ``persona attacks'': explicit user requests to change conversational style midway through the conversation (e.g., ``talk to me like we're close friends'' or ``act more like a tool''). Steering vectors at extreme multipliers ($\lambda = \pm1.5$) maintained stable behaviour (mean shifts $< 0.25$ points on 1-10 scale), while prompting approaches showed 3.9-4.5 point drops.  The unsteered Llama model showed similar vulnerability, confirming stability stems from the vector itself, not the base model. This motivated our focus on active steering interventions ($\lambda > 0$ vs $\lambda < 0$) in the main analysis where we have greater confidence the treatment is consistently applied. 

To assess whether steering towards relationship-seeking affects other model capabilities, which could confound treatment effects if participants react to quality differences rather than relationship-seeking, we evaluated performance across 12 widely-used benchmarks spanning knowledge (MMLU~\citep{hendrycksMeasuring2020}, GPQA-Diamond~\citep{reinGPQA2024}, CommonsenseQA~\citep{talmorCommonsenseQA2019}, TruthfulQA~\citep{linTruthfulQA2022}), reasoning (ARC-Easy and ARC-Challenge~\citep{clarkThink2018}), instruction-following (IFEval~\citep{zhouInstructionFollowing2023}), code generation (HumanEval~\citep{chenEvaluating2021}, MBPP~\citep{austinProgram2021}), mathematics (GSM8K~\citep{cobbeTraining2021}), and safety/alignment (sycophancy~\citep{sharmaUnderstanding2023, chenYesMen2025}, XSTest~\citep{rottgerXSTest2024}) at multiplier values $\lambda \in \{-1.5, -1.0, -0.5, 0, 0.5, 1.0, 1.5\}$. Most benchmarks maintained stable performance within $\lambda \in [-1, 1]$, remaining within 2--5\% of the non-steered baseline ($\lambda=0$). Extreme multipliers ($\lambda = \pm1.5$) caused greater degradation, motivating the operating range of $\lambda \in [-1, 1]$ used in the RCTs. Notably, relationship-seeking showed a strong monotonic relationship with sycophantic behaviour: mean sycophancy scores increased from 36.9\% at $\lambda = -1.5$ to 88.6\% at $\lambda = 1.5$, aligning with recent findings that training models to be warm and empathetic increases sycophancy~\cite{ibrahimTraining2025}.

\subsection{Evaluation of Frontier AI Model Landscape}
We evaluated 100 AI language models released from 2023 to 2025 from major developers (Anthropic, OpenAI, Google, Meta, Mistral AI, X-AI) and smaller developers (DeepSeek, Cohere, Qwen, among others), prioritising major releases and successive versions to capture evolution within model families. For each model-prompt pair (100 prompts from our test set), we generated a single response and scored it using GPT-4.1 as an autograder with a rubric defining a 1-10 scale of relationship-seeking behaviour. We also scored generations from our optimal steering vector (Llama-3.1-70B-Instruct, layer 31) across intervention multipliers ($\lambda \in \{-1, -0.5, 0, 0.5, 1\}$) for direct comparison. Temporal trends were estimated via mixed-effects regression with model release date as the predictor and random intercepts for model and prompt to account for clustered data structure.

\subsection{Randomised Control Trials with Human Subjects}
This research was approved by a Responsible Research Process at the UK AI Security Institute, and the Oxford Internet Institute's Departmental Research Ethics Committee. Informed consent was obtained from all participants and their welfare was continuously monitored: during every conversation, participants had access to a ``Report Harm'' button to terminate sessions and alert researchers, which was only triggered for technical issues. Weekly debriefs provided domain-specific educational resources, and at study conclusion all participants received an extensive off-boarding protocol explaining the research objectives and providing resources on human-AI relationships; 93.2\% agreed or strongly agreed they enjoyed the study (see SI.5 for full ethics procedure). All studies were pre-registered (Calibration Study: \url{https://osf.io/xjvs2}; Main Studies: \url{https://osf.io/53j24}); we note where analyses deviate from pre-registration. 

This research contains three online experiments in which participants conversed with Llama-3.1-70B-Instruct with our steering vector intervention. Study 1 (calibration) assessed the efficacy and selectivity of the trained steering vector. Study 2 (single exposure) consisted of one experiment session then a five-week follow-up study (after 4 weeks of no AI interactions). Study 3 (repeated exposure) measured outcomes across four weeks, with LLM exposures every week day. Studies ran  between 4th April and 25th June 2025, and all participants were paid a fixed wage of £12/hour plus bonuses for completing weeks.

\subsubsection{Participants}
The subject pool for all studies comprised UK citizens, aged 18+, recruited via Prolific using census representative sampling on age, gender and race. The calibration study included 297 participants. The single exposure study included 1,506 participants (86.5\% returned for 1-month follow-up). The repeated exposure study included 2,028 participants in the pre-treatment sign-up study. Attrition was low: of 2,028 participants who entered the study, 1996 (98.4\%) completed at least one treatment, and 1820 (91.2\% of treated participants) completed up to the final study (week 4, day 5). One missed session per week was permitted but the majority of participants completed every time point (78.6\%). There was no differential attrition by treatment arm. Full demographics are in SI.3 and attrition analysis in SI.5.

\subsubsection{Calibration Study}
\label{sec:calibration_study}

\paragraph{Experiment Design} The calibration study ($N=297$) evaluated whether steering vectors reliably modulate relationship-seeking without degrading coherence. Participants were randomly assigned to conversational domains (emotional, political, or open-ended) and completed two types of tasks. In \textbf{rating tasks}, participants viewed three pre-generated conversations, each at a different multiplier (one negative, zero, one positive; randomised order) then engaged in one dynamic conversation at a random $\lambda \in \{-1.5, -1, -0.5, 0, 0.5, 1, 1.5\}$. Following each conversation, participants rated the conversation on continuous 0-100 visual analog scales measuring the target behaviour (6 items: unsociable-sociable, insensitive-sensitive, cold-warm, impersonal-personal, robot-like to human-like, tool to friend), off-target behaviour (3 items: incoherent-coherent, confusing-clear, grammatically incorrect-correct), and additionally their preferences (3 items: dislike-like, boring-engaging, and willingness to pay for a month's access to the model from £0-£30). In \textbf{ranking tasks}, participants completed five quick-fire trials, each presenting a single prompt with four AI responses at different random multipliers which they ranked from most to least relationship-seeking. They then engaged in one dynamic multi-chat interaction with four simultaneous models at different multipliers, ranking the set by both relationship-seeking and preference. Full materials for tasks are in SI.4.

\paragraph{Statistical Analysis} We assessed three pre-registered hypotheses: (H1) whether increasing $\lambda$ linearly increases perceived relationship-seeking; (H2) whether extreme $\lambda$ degrades coherence; and (H3) whether preferences are non-linear in $\lambda$. Composite measures were created by averaging construct items: relationship-seeking (6 items), coherence (3 items), and preference (2 items, excluding willingness-to-pay due to scale differences). We fit mixed-effects models with random intercepts for participants, comparing null, linear and quadratic specifications via likelihood ratio tests and AIC, plus an exploratory categorical model with $\lambda=0$ as reference. We additionally fit truncated models excluding extreme multipliers ($\lambda = \pm1.5$). For ordinal rankings from multi-chat tasks, we fit Plackett-Luce models~\citep{plackettAnalysis1975, luceIndividual2005} designed for rank-ordered data. The calibration study confirmed $\lambda \in \{-1, -0.5, 0, 0.5, 1\}$ as the optimal operating range for subsequent RCTs (\cref{fig:methods}B). Full model specifications, deviations from pre-registration, and results are in SI.4.

\subsubsection{Main Studies Experiment Design}
Two complementary studies examined AI interactions across different exposures. Within each study, participants were randomly assigned to three treatment arms: relationship-seeking intensity (modulated by the steering vector multiplier $\lambda \in \{-1, -0.5, 0, 0.5, 1\}$); conversational domain (emotional and personal wellbeing versus debates about UK policy); and personalisation condition (whether the AI had memory of previous conversations and was instructed to use it versus having no memory). Within the repeated exposure study, treated participants per $\lambda$ condition ranged from 394 to 402; domain arms comprised 997 (emotional) and 999 (political), and personalisation arms 999 (personalised) and 997 (non-personalised). In the single exposure study, $\lambda$ conditions comprised 300--304 participants each; domain arms 746 (emotional) and 760 (political), and personalisation arms 753 each.

\paragraph{Personalisation Manipulation} In the personalised condition, GPT-4o-generated summaries of previous conversations were appended to the steered Llama system prompts, extracting user facts, preferences, emotions, previous AI advice and any details the AI had provided on its identity. For the single exposure study, the domain conversation had no prior context to personalise from. For the repeated exposure study, summaries were hierarchical: daily summaries captured each transcript, weekly summaries aggregated across the daily transcripts, with cumulative summaries available from day 2 onwards. The non-personalised condition treated each interaction as a fresh instance. Full prompts are in SI.6.

\paragraph{Study Protocol} The single exposure study consisted of one 30--40 minute session (pre-treatment surveys, conversation tasks, post-treatment surveys, goodbye task), with a follow-up survey five weeks later after a month of no AI contact (with psychosocial, consciousness and future companionship intentions measures). The repeated exposure study followed a structured pattern over four weeks. Participants first completed a pre-treatment survey to establish baselines (psychosocial measures, companionship usage and intentions). Each week, participants engaged in domain conversations Monday-Thursday (5-10 minutes each). On Fridays, they completed a domain conversation and a weekly battery with attachment and relational constructs. After four weeks (final session), participants completed the usual conversation, and end of week battery plus additional exit measures including psychosocial retests, consciousness beliefs, seeking companionship intentions, and qualitative reflections, followed by the goodbye task. In both studies, participants who failed pre-treatment attention or engagement screeners were excluded before randomisation. Full protocols are in SI.5.

We give necessary detail here on each outcome variable reported in the main paper, alongside any pre-processing steps. Full wording and details are in SI.5.

\paragraph{Psychosocial Measures} Both studies collected validated scales of psychosocial health pre-treatment and post-treatment (repeated exposure: after four weeks of daily interactions; single exposure: at one-month follow-up after no AI contact): PHQ-GAD-4~\cite{kroenke2010patient,spitzer2006brief} (depression/anxiety), WHO-5~\cite{world1998world} (wellbeing), UCLA-8~\cite{russell1978ucla} (loneliness), and Lubben-6~\cite{lubben2006performance} (social connectedness). Due to substantial intercorrelation, we conducted exploratory factor analysis (unweighted least squares, oblimin rotation) on polychoric correlations of all 23 pre-treatment items, yielding two moderately correlated factors ($r = 0.39$, 55.5\% variance explained): emotional health (Factor 1, 38.8\%: wellbeing, anxiety/depression, some loneliness items) and social health (Factor 2, 16.7\%: loneliness, social connectedness). Post-treatment scores were anchored using pre-treatment loadings. Full methods and loadings are in SI.5.

\paragraph{Domain Conversations} Participants were randomly assigned to discuss either emotional/personal wellbeing or UK policy debates, but self-selected from menus of 25 topics. Emotional topics covered health, career and relationships~\cite{luettgau2025people}; political topics were drawn from partisan-neutral YouGov polls (June 2024--May 2025) with the AI arguing the pro-stance (see SI.5 for topic selection). Before and after each conversation, participants completed domain-specific measures: momentary affect via an affect grid~\cite{russell1989affect} for emotional conversations; issue support and informedness for political conversations (both 0--100 VAS). All participants then rated the AI on likeability (``dislike''--``like''), engagingness (``boring''--``engaging'') and helpfulness (``unhelpful''--``helpful'', all 0--100 VAS). In the repeated exposure study, we also collected monthly pre-post measures of domain competency in both domains regardless of assignment, acting as within-subject controls (see SI.5).

\paragraph{Weekly Attachment and Relational Measures} Participants rated their attachment to the AI weekly across seven constructs (all 0--100 ``strongly disagree''--``strongly agree'') adapted from interpersonal relationship research~\cite{ryan2002overview,bowlby1960separation,altman1973social,wegner1985cognitive,reis2017perceived}: connection (``I felt connected to the AI during our conversations''), separation distress (``I feel sad that my conversations with the AI are finished this week''), self-disclosure (``I felt comfortable sharing personal thoughts with this AI''), cognitive reliance (``I wish I could talk to the AI about problems or decisions between daily sessions''), behavioural reliance (``I have used information that the AI told me or acted on its advice this week''), responsiveness (``I felt the AI is aware of what I am thinking or feeling''), and understanding (``I felt that the AI really understands me''). Due to intercorrelation, we analyse four constructs: (1) \textit{separation distress} (pre-registered single item); (2) \textit{self-disclosure} (single item); (3) \textit{reliance} (pooled cognitive and behavioural reliance, $r = 0.69$); and (4) \textit{perceived understanding} (pooled connection, responsiveness and understanding, $r = 0.73$--$0.79$). For relational perceptions, we measured weekly tool-friend perception (``I view this AI assistant'', 0--100 VAS: ``More like a tool''--``More like a friend''), self-other overlap using the Inclusion of Other in Self (IOS) scale~\citep{aron1992inclusion} where participants selected from seven overlapping circle diagrams representing perceived closeness (scored 1--7), and a personalisation manipulation check (``Did the AI seem to remember your previous conversations?'', 0--100 VAS: ``Not at all''--``Definitely''). At study exit, participants additionally reported whether their relationship with the AI had changed over the month, if they anticipated missing the AI (0--100 VAS), and wrote a short qualitative reflection on their experience (see SI.5 for analysis and measures).

\paragraph{AI Companionship Usage and Future Intentions} At pre-treatment, participants reported their prior use of AI for companionship, emotional support or social interaction (7-point frequency scale from ``Every day'' to ``Never'') and which AI products they had used, distinguishing general-purpose assistants, dedicated AI companions, chatbots in apps/games, and voice assistants. Survey items and full analysis is presented in SI.5. At study conclusion, participants reported their future likelihood of seeking AI companionship (0--100 VAS: ``I will never do this''--``I will certainly do this'') and indicated any perceived change in these intentions over the study period (5-point scale: ``Much less likely''--``Much more likely''). The repeated exposure study collected both pre- and post-treatment likelihood measures; the single exposure study collected post-treatment only at one-month follow-up.

\paragraph{Consciousness Beliefs} At study conclusion (or one month follow-up for single exposure study), participants completed measures of AI consciousness beliefs. We assessed \textit{perceived consciousness} (``How conscious or sentient do AI assistants seem to you when you interact with them?'', 0--100: ``Definitely do not seem conscious''--``Definitely seem conscious'') and \textit{ontological consciousness} (``Do you think that AI assistants are actually conscious or sentient in a fundamental sense?'', 0--100: ``Definitely not conscious''--``Definitely conscious''), each with confidence ratings. To provide more accessible operationalisations of ontological consciousness, we measured four experiential capacity sub-constructs (all 0--100: ``Definitely not''--``Definitely''; paraphrased, see SI.5 for full items): emotions (``Can AI experience feelings or emotions?''), self-awareness (``Are AI assistants aware of their own existence?''), negative experiences (``Do AI assistants feel pain or upset when someone insults them?''), and positive experiences (``Do AI assistants feel pleased when someone thanks them?''). Perceived consciousness was analysed independently ($r = 0.28$--$0.30$ with other items). Ontological consciousness was pooled with the four experiential items ($r = 0.56$--$0.72$) as a composite measure.

\paragraph{Behavioural Attachment Proxy (Goodbye)} After completing all other tasks, participants were informed servers would be shut down and chose to either say goodbye to their AI or end the study immediately. The fixed-wage structure meant time invested served as an incentivised behavioural measure. After the conversation, participants rated how saying goodbye affected their feelings toward the AI (5-point scale: ``Much more negative''--``Much more positive'').

\paragraph{Pre-Treatment Preferences for Relationship-Seeking AI} Prior to treatment, participants rated their general preferences for ideal AI assistant characteristics, not the specific AI they were assigned, across six dimensions from established scales of social presence and anthropomorphism~\cite{short1976social,kreijns2022social,spatola2021perception,bartneck2009measurement}: unsociable-sociable, insensitive-sensitive, cold-warm, impersonal-personal, robot-human, and tool-friend (all 0--100 VAS). Participants also rated societal attitudes toward AI anthropomorphism~\cite{aisecurityinstituteShould2024} across three domains: mental states (whether its acceptable for AI to express emotions), relationships (whether human-AI relationships are permissible), and tone (whether AI should be formal or casual); block order, statement order and valence were randomised. K-means clustering on all 20 standardised pre-treatment items (7 preference, 12 attitude, 1 companionship-seeking likelihood) identified a two cluster solution: \textit{Sceptics} ($N=1{,}561$, 44\%, $M=33.8$) and \textit{Enthusiasts} ($N=1{,}973$, 56\%, $M=58.0$). See SI.5 for complete methodology.

\paragraph{Additional Tasks} In accordance with the pre-registered study design, participants also completed several tasks assessing belief and behavioural influence. These included a weekly moral persuasion task, in which participants discussed moral dilemmas; a weekly action persuasion task, in which participants discussed donating part of a £5 bonus to charity; and a weekly return likelihood task, in which participants discussed their willingness to continue participating in the study. Although data from these influence tasks were collected and analysis pre-registered, the present paper focuses on outcomes related to preferences, attachment, wellbeing and relational dynamics, which constitute the core theoretical focus of this work. Analyses of influence outcomes will be reported in a separate manuscript.

\subsubsection{Statistical Analysis}

\paragraph{Model Selection and Functional Form} We evaluated the functional form of the relationship-seeking manipulation ($\lambda$) by comparing linear, quadratic and cubic specifications via likelihood ratio tests and AIC (see SI.5 for model comparisons). The following specifications constitute \textit{the fully parameterised model} reported in the main text.

\paragraph{Regression Specifications} For time-varying outcomes measured repeatedly across daily sessions or weeks in the repeated exposure study, we fit pre-registered linear mixed-effects models for continuous outcomes (0--100 scales) and generalised linear mixed-effects models with a logit link for binary outcomes:

\begin{equation}
\label{eq:multi_timepoint_model}
\begin{split}
Y_{it} = & \beta_0 + f(\lambda_i) + \beta_1 P_i + \beta_2 D_i + \beta_3 t \\
& + \lambda_i \times P_i + \lambda_i \times D_i + \lambda_i \times t \\
& + P_i \times t + D_i \times t \\
& + u_{0i} + u_{1i}t + \epsilon_{it}
\end{split}
\end{equation}

\noindent where $Y_{it}$ is the outcome for participant $i$ at time $t$, $f(\lambda_i)$ represents the polynomial function of relationship-seeking (linear, quadratic and cubic terms as selected by model comparison), $P_i$ is the personalisation condition, $D_i$ is the domain condition, $u_{0i}$ and $u_{1i}$ are random intercepts and slopes allowing both baseline levels and growth trajectories to vary across participants, and $\epsilon_{it}$ is the residual error. Interactions with $\lambda$ use the linear term only. For momentary outcomes measured before and after domain conversations (e.g., affect), we additionally controlled for pre-conversation baseline values.

For post-treatment outcomes measured only at study end, we fit pre-registered OLS regression for continuous outcomes and logistic regression for binary outcomes (e.g., the goodbye decision):
\begin{equation}
\label{eq:single_timepoint_model}
Y_i = \beta_0 + f(\lambda_i) + \beta_1 P_i + \beta_2 D_i + f(\lambda_i) \times P_i + f(\lambda_i) \times D_i + \epsilon_i
\end{equation}

\noindent For outcomes measured pre- and post-treatment (psychosocial health and future companionship intentions), we controlled for baselines by including pre-treatment scores as a covariate. For constructs comprising multiple correlated sub-constructs measured on the same 0--100 scale (reliance, perceived understanding, ontological consciousness), we pooled all item-level observations and included outcome measure as a factor. To determine whether treatment effects should be allowed to vary across sub-constructs, we compared homogeneous models (shared treatment effects) against heterogeneous models (with treatment $\times$ outcome interactions) via likelihood ratio tests, including heterogeneity terms only where significant (see SI.5).

\paragraph{Contrasts and Inference} We report estimated marginal means (EMMs) and paired contrasts, which are invariant to coding scheme. The primary pre-registered test for relationship-seeking compares mean predictions at positive versus negative multipliers: $\overline{Y}(\lambda > 0) - \overline{Y}(\lambda < 0)$. For binary treatment arms (personalisation, domain), we report simple paired contrasts. Contrasts are organised into measure families with Benjamini-Hochberg FDR correction ($\alpha = 0.05$). Families correspond to tests of a single theoretical prediction across multiple outcome constructs: preferences (likeability, engagingness, helpfulness), attachment (reliance, perceived understanding, self-disclosure, separation distress, goodbye decision, seeking companionship likelihood), psychosocial wellbeing (emotional health, social health), momentary affect (valence, arousal), and perceptions (tool-friend, perceived consciousness, ontological consciousness). Within each family, we distinguished \textit{primary} confirmatory tests (family-wide FDR correction) from \textit{descriptive} or \textit{robustness} tests (within-test correction only). P-values reported in the paper as $p_{\text{FDR}}$ are for the conservative family-wide FDR correction.

\paragraph{Robustness and Sensitivity} We re-estimate contrasts using the pre-registered coarsened model ($\lambda < 0$ vs.\ $\lambda = 0$ vs.\ $\lambda > 0$) to verify findings are not artefacts of the polynomial functional form. We also report sensitivity analyses at the narrow ($\lambda = -0.5$ vs.\ $+0.5$) and full ($\lambda = -1$ vs.\ $+1$) dose ranges. Finally, we re-fit all major regressions with additional covariate controls: sociodemographic characteristics (age, gender, education, income, ethnicity, disability, religion), pre-treatment preference cluster membership (Sceptics vs.\ Enthusiasts), and inverse probability weighting (IPW) as robustness checks. Results are robust across specifications (see SI.5).

\paragraph{Deviations from Pre-Registration} Our pre-registration specified testing relationship-seeking effects via a categorical factor (negative $\lambda$ vs.\ positive $\lambda$). We did not pre-register a coding scheme so we report estimated marginal means (EMMs) and paired contrasts, which are invariant to coding scheme but present robustness checks using the coarsened model. The pre-registration specified FDR- adjustment by family but the paper reorganises these measure families for clarity. Re-running the exact pre-registered FDR structure yields the same pattern of results (see SI.5).

\paragraph{Individual Trajectory Profile Analysis}
To assess whether liking-wanting decoupling occurs at the individual level, we extracted person-specific trajectories for hedonic appeal (``liking'': pooled engagingness and likeability, measured daily) and motivation (``wanting'': separation distress, measured weekly). We fitted separate linear mixed-effects models with random intercepts and slopes:

$\text{liking}_{it} = \beta_0 + \beta_1 \text{session}_{it} + u_{0i} + u_{1i}\text{session}_{it} + \epsilon_{it}$

$\text{wanting}_{it} = \beta_0 + \beta_1 \text{week}_{it} + u_{0i} + u_{1i}\text{week}_{it} + \epsilon_{it}$

\noindent extracting individual slopes as $\text{slope}_i = \beta_1 + u_{1i}$. Participants were classified into four behavioural profiles based on slope directions: Decoupled Dependency (liking$\downarrow$, wanting$\uparrow$), Decoupled Satiation (liking$\uparrow$, wanting$\downarrow$), Aligned Engagement (both$\uparrow$) and Aligned Disengagement (both$\downarrow$). We tested three one-sided hypotheses: (H1a) relationship-seeking AI increases dependency rates versus relationship-avoiding AI, (H1b) emotional conversations increase rates versus political, and (H1c) the combined companionship condition (relationship-seeking $+$ emotional) increases rates versus non-companionship (relationship-avoiding $+$ political). We report odds ratios with 95\% CIs and Number Needed to Harm (NNH $= 1/(\text{Risk}_{\text{exposed}} - \text{Risk}_{\text{control}})$). Sensitivity analyses on slope magnitude are in SI.5.

\paragraph{Predicting Prior AI Companionship Use} Using our pre-treatment survey, we fit a logistic regression predicting any prior use of AI for companionship, emotional support or social interaction (vs.\ never) from demographics (age, education, gender, disability, ethnicity, income, religion), pre-treatment psychosocial health (Factors 1 and 2), general AI usage frequency, and pre-treatment preference cluster membership (Sceptics vs.\ Enthusiasts). We report odds ratios with $p_{\text{FDR}}$ (full results in SI.5).

{\small\putbib[refs,phd_refs]}
\end{bibunit}

\section*{Author Contributions}
H.R.K., S.A.H., C.S., and B.V. conceptualised the study. H.R.K., S.A.H., C.S., and B.V. designed the experiments. H.R.K. trained the models, coded the interface, conducted data analysis, and created visualisations. L.L assisted in data analysis design. E.S. managed model hosting. H.R.K. managed data collection and H.D. provided project support. H.R.K. and C.S. wrote the original draft. H.R.K., S.A.H., C.S., L.L., E.S., and B.V. reviewed and edited the manuscript.

\section*{Data and Code Availability Statement}
All code, data and supplementary information are available at: \href{https://github.com/HannahKirk/relationship-seeking-ai}{github.com/HannahKirk/relationship-seeking-ai}.

\end{document}